\documentclass[
preprint,
preprintnumbers,
tightenlines,
superscriptaddress,
amsmath,
amssymb,
noshowkeys,
nofootinbib,
floatfix,
]{revtex4-2}
\usepackage{graphicx}          
\usepackage{subfigure}         
\usepackage{dcolumn}           
\usepackage{capt-of}           
\usepackage{makecell}          
\usepackage{bbm}               
\usepackage{bm}                
\usepackage{time}              
\usepackage[mathlines]{lineno} 
\usepackage{stackrel}          
\usepackage{enumitem}          
\usepackage{physics}           
\usepackage[mathscr]{eucal}    
\usepackage{hyperref}          
\hypersetup{
    pdfnewwindow=true,         
    colorlinks=true,           
    linkcolor=red,             
    citecolor=blue,            
    filecolor=green,           
    urlcolor=cyan              
}
\usepackage{amsthm}
\theoremstyle{plain}                           

\theoremstyle{definition}

\theoremstyle{remark}

\newcommand{\tworule}{\vspace{\baselineskip}\noindent\hrule\vskip\doublerulesep\hrule}

\newcommand{\rmi}{{\rm i}}                     
\newcommand{\Partial}[4]
   {\Bigl ( \frac{\partial #1 }{\partial #2 } \Bigr )_{\! #3, #4 }}
\newcommand{\Tproduct}[1]%
   {\ensuremath{\mathcal{T} \{ \, #1 \, \} } }
\begin{document} 
\newcommand{\pdhfrac}[2]{\mathchoice{\frac{#1}{#2}}{#1/#2}{#1/#2}{#1/#2}}
\newcommand{\fdd}[2]{\pdhfrac{\mathrm{d}#1}{\mathrm{d}#2}}
\newcommand{\sdd}[2]{\pdhfrac{\mathrm{d}^2#1}{\mathrm{d}#2^2}}
%
%
%
\newcommand{\be}{\begin{equation}}
\newcommand{\ee}{\end{equation}}
\newcommand{\bea}{\begin{eqnarray}}
\newcommand{\eea}{\end{eqnarray}}
\newcommand{\tchi} {{\tilde \chi}}
\newcommand{\tA} {{\tilde A}}
\newcommand{\tp} {{\tilde p}}
\newcommand{\tir} {{\tilde r}}
\newcommand{\bq}{\begin{equation}}
\newcommand{\ba}{\begin{eqnarray}}
\newcommand{\eq}{\end{equation}}

%

%
\preprint{LA-UR-26}
\title{Nonlinear Dirac Equations With All Possible Combinations of 
Scalar-Scalar, Vector-Vector and Pseudoscalar-Pseudoscalar Interactions 
}
\author{Avinash Khare}
\email[Email: ]{avinashkhare45@gmail.com} 
\affiliation{
   Physics Department, 
   Savitribai Phule Pune University, 
   Pune 411007, India} 
\author{Fred Cooper}
\email[Email: ]{cooper@santafe.edu}
\affiliation{
   Santa Fe Institute,
   1399 Hyde Park Road,
   Santa Fe, NM 87501, USA}
\affiliation{
   Center for Nonlinear Studies and Theoretical Division, 
   Los Alamos National Laboratory, 
   Los Alamos, NM 87545, USA}
\author{John F. Dawson}
\email[Email: ]{john.dawson@unh.edu}
\affiliation{
   Department of Physics,
   University of New Hampshire,
   Durham, NH 03824, USA}
\author{Avadh Saxena} 
\email[Email: ]{avadh@lanl.gov} 
\affiliation{
   Center for Nonlinear Studies and Theoretical Division, 
   Los Alamos National Laboratory, 
   Los Alamos, NM 87545, USA}
\date{\today, \now \ PST}
\begin{abstract}
We obtain exact solutions of the nonlinear Dirac (NLD) equation in $1+1$ dimensions 
of the form $\psi(x,t) = e^{-i\omega t} \psi(x)$ for all possible combinations 
of scalar-scalar (SS), vector-vector (VV) and pseudoscalar-pseudoscalar (PP) 
interactions at nonlinearity parameter $\kappa = 1$, which corresponds to the 
most general  4-Fermion interactions. We look for solutions where $0 < \omega < m$ 
where $\omega, m$ are frequency and mass respectively. Remarkably, we show that
at $\kappa = 1$, all the NLD models with SS, VV and PP interactions can be 
reduced to suitable combinations of SS and VV interactions and hence all 
solitary waves in these models must also be bound states. We show that
two of the ten possible combinations are equivalent to $SS -(1/p) VV$ 
interactions while two other combinations are equivalent to $VV -(1/q) SS$ 
interactions for which we already know the results. For the
remaining six combinations of SS, VV and PP, we show that depending on the 
parameters, whereas in some cases entire range of $0 < \omega < m$ is allowed, 
in some other cases $\omega$ is restricted to a smaller range 
$\omega_{min} < \omega < m$. Further, depending on the parameters, while the solitary
 waves in some cases have only one hump, in other cases the solitary waves have 
 both single and double hump depending on the value of $\omega$. We 
compute the charge $Q$ and the energy $E$ for each of the solitary wave 
solutions and show that while $E$ and $Q$ depend on coupling constant $g$, 
their ratio $E/Q$ is independent of $g$.
When $\kappa = 1$, all the NLD models with SS, VV and PP interactions can be reduced
to suitable combinations of SS and VV interactions.  
As a result all solitary waves
in these models  are  bound states with $E/Q<m$.  
\end{abstract}
\maketitle
\tworule
\section{\label{s:Intro}Introduction}

The relativistic description of fermions by the Dirac equation is an important aspect  
of modern gauge theories. Subsequent to the seminal paper of Dirac and others, 
the nonlinear Dirac (NLD) equation with scalar-scalar (S-S) as well as vector-vector 
(V-V) interactions was introduced by Ivanenko \cite{Iv1} and Thirring \cite{R3},  
respectively. Initially, these models were extensively applied in particle physics; 
for instance to describe extended nucleons \cite{R4} or to understand quark 
confinement \cite{R5}. Later, these models found diverse applications such as 
in optical gratings \cite{gratings}. Additional applications of the Dirac equation 
include the light propagation in honeycomb photo-refractive lattices, e.g. photonic 
graphene \cite{R10}, conical diffraction in this kind of structures \cite{R11}, 
spin-orbit coupled Bose-Einstein condensates \cite{R12} besides in the area of 
conducting polymers \cite{bc}. 

The explicit solutions of the $1+1$ dimensional Soler model as well as Thirring
model were known since 1974 \cite{lee}. In 2010 these one-dimensional models
were generalized to arbitrary nonlinear parameter $\kappa$ and explicit 
solitary wave solutions were obtained for arbitrary $\kappa > 0$ \cite{fred}. 
Further, it was shown that whereas solitary wave bound states exist for any 
$\kappa > 0$ in the scalar-scalar interaction case (SS), in the vector-vector
case it was shown that these bound states only exist if 
$\kappa < \kappa_c$. Recently, Alexeeva, Barashenkov and Saxena \cite{abs} 
introduced a novel model at $\kappa = 1$ with an admixture of both the SS and the 
VV interactions. Recently we \cite{ak} generalized this model to a continuous 
two-parameter family of models (which we denote in what follows as the 
gABS model) characterized by the nonlinearity parameter $\kappa > 0$ as 
well as a parameter $p > 1$ characterizing the strength of the VV interaction 
vis a vis the SS interaction. In that paper we showed that 
the solitary wave bound states exist in the entire $\kappa$-$p$ plane. 
We also recently considered the complementary gABS model in which the SS  
interaction term has negative sign while the VV interaction term has the 
positive sign \cite{cgabs}.  That is the complementary NLD model characterized 
by $VV +(1/q) SS$ with $q < -1$. We showed that unlike the $SS+(1/p)VV$ model 
with $p < -1$, in this model solitary wave bound states exist only if 
$\kappa < \kappa_c$ which is approximately $2.5$ \cite{cgabs}. 
More recently we have investigated the solitary waves that exist when the 
interactions are of the form $VV+SS$ and mapped out the regimes of bound state 
solutions as well as when there are two-humped vs. one-humped solutions 
\cite{fred7}. 

It is worth noting that the most general NLD models can have an admixture of not
only the scalar-scalar and the vector-vector but also the pseudoscalar-pseudoscalar 
(PP) interactions. In this context it is worth remembering that in $1+1$ dimensions,
pseudovector-pseudovector interactions are equivalent to the VV interactions 
and that tensor interactions can be related to pseudoscalar ones 
\cite{shaovsp}. It is then natural to consider the most general NLD models
including all three, i.e. SS, VV and PP interactions. The purpose of this paper
is to take a first step in that direction by considering NLD models with all 
possible combinations of SS, VV and PP interactions at $\kappa = 1$. It turns 
out that there are ten such possible combinations.

Remarkably, we show that at 
$\kappa = 1$, all the NLD models with SS, VV and PP interactions can be reduced
to suitable combinations of SS and VV interactions. Since for all the 
combinations of the NLD models with the SS and the VV interactions, we know 
that at $\kappa = 1$, all the solitary waves are solitary wave bound states 
(i.e. for all of them $E/Q < m$, the mass) it then follows that the most
general NLD model with SS, VV and PP interaction also has solitary wave bound
states at $\kappa = 1$.  We show that
two of the ten possible combinations are equivalent to $SS +(1/p) VV$ 
interactions, while two other combinations are equivalent to 
$VV +(1/q) SS$ interactions with $p, q < -1$ for which we already know the 
results \cite{fred7}. For the remaining six combinations of SS, VV and PP, we 
compute the charge $Q$ and the energy $E$ for each of the solitary wave 
solutions and show that while $E$ and $Q$ depend on the coupling constant $g$, 
their ratio $E/Q$ is independent of $g$. For various values of $q,p,r$ and 
$\kappa =1$  we find the transition from double to single hump solitary waves 
and we also determine the ratio $E/Q$ as a function of $\omega/m$  
where $E$ is the energy and $Q$ is the charge of the solitary wave.  
We will parameterize the various possible combinations of SS, VV, PP 
interactions using the parameters $q$, $p$, $r$ and an interaction Lagrangian
\be \label{1a} 
L_I = \frac{g^2}{q(\kappa+1)}(\bar{\psi} \psi)^{\kappa+1}+\frac{g^2}
{r(\kappa+1)}[(\rmi \bar{\psi} \gamma^{5} \psi )
(\rmi  \bar{\psi} \gamma^5 \psi) ]^{(\kappa+1)/2} 
+\frac{g^2}{p(\kappa+1)}[\bar{\psi} \gamma_{\mu}\psi 
\bar{\psi} \gamma^{\mu} \psi]^{(\kappa+1)/2}\,.
\ee
At  $\kappa=1$ the interaction Lagrangian reduces to
\be\label{2}
 L_I = \frac{g^2}{2q}(\bar{\psi} \psi)^{2}+\frac{g^2}
{2r}[\bar{\psi}\rmi  \gamma^{5}\psi 
\bar{\psi} \rmi\gamma^5 \psi]
+\frac{g^2}{2p}[\bar{\psi} \gamma_{\mu}\psi 
\bar{\psi} \gamma^{\mu} \psi]\,.
\ee
A note of caution here. It may appear that one is considering the most general 
model with SS, VV and PP interactions at $\kappa = 1$ having three arbitrary
parameters $q,p,r$. However, this is not so. It turns out that in all the ten
combinations of SS, VV and PP interactions one either has $q = 1$ or 
$p = 1$ so that effectively one is considering two-parameter family of various
combinations of SS, VV and PP interactions.

The plan of the paper is the following. In Sec. II we obtain the rest-frame
solitary wave solutions of the form $\psi(x,t) = e^{-i\omega t} \psi(x)$ 
which are functions of the rest-frame frequency $\omega$ as well as $m, \kappa$
and $g$. In this paper we are interested in looking for solutions for which
$0 < \omega < m$. In Sec. III we show that at $\kappa = 1$, all allowable 
combinations of NLD models with interaction including SS, VV and PP are 
equivalent to suitable combinations of SS and VV interactions. In Sec. IV we 
discuss the various properties of the solitary wave solutions and determine the
regime of single hump and double hump solutions as a function of $\omega/m$ and
the parameters $p,r,q$. Further, we calculate the total energy $E$ as well as 
the charge $Q$ and show that while $E$ and $Q$ depend on the coupling constant 
$g$, their ratio $E/Q$ is independent of $g$. In Sec. V we summarize the main 
results obtained in this paper.

\section{ Nonlinear Dirac Equation With SS, VV, and PP Interactions}

The Lagrangian density for the  SS+VV+P-P model is of the form:
\be\label{1}
L = \bar{\psi} ( i \gamma^{\mu} \partial_{\mu} -m) \psi + L_I\,,
\ee
where $L_I$ is as given in  Eq. (\ref{2}). 

\subsection{ Solitary Wave Solutions}

We will follow the notations of our 2010 paper \cite{fred}.  It turns out 
that the several basic results given there are also valid for our case.
We are interested in finding the solitary wave solutions of the generalized  
model characterized by the Lagrangian as given by Eqs. (\ref{1}). 
The corresponding field equation is
\bea\label{3a}
( i \gamma^{\mu} \partial_{\mu} - m) \psi + \frac{g^2}{q} (\bar{\psi} \psi)\psi
+\frac{g^2}{r} (\bar{\psi}\rmi \gamma^5 \psi)\rmi \gamma^{5} \psi
+\frac{g^2}{p} (\bar{\psi}\gamma_{\mu} \psi) \gamma^{\mu} \psi
= 0\,.
\eea
We  will use the same  representation of $\gamma$ matrices as used in \cite{fred}:
\be\label{4}
\gamma^{0} = \sigma_3 = 
\begin{pmatrix}
 1 & 0\\
 0 & -1 
 \end{pmatrix}\,~~ , 
\gamma^{1} = i \sigma_1 = 
i \begin{pmatrix} 0 & 1 \\ 
1 & 0 
\end{pmatrix} 
\ee

and
\be \label{gam5}
\gamma^{5} = \gamma^0 \gamma^1= - \sigma_2 = 
\begin{pmatrix}
 0 & \rmi \\
 - \rmi & 0 
 \end{pmatrix}\,, 
 \ee
where $\sigma_i$ are the Pauli matrices. The gamma 
matrices then obey the 
anti-commutation relation, $[\gamma^{\mu},\gamma^{\nu}]_{+} = 2 g^{\mu\nu}$. 

We look for solutions (in the solitary wave rest-frame) of the form,
\be\label{5}
\psi(x,t) = (u(x), v(x)) \, e^{-i \omega t}
= R(x) (\cos(\theta(x)), \sin(\theta(x))) \, e^{-i \omega t} \,. 
\ee
From current conservation
\be\label{6}
\partial_{\mu} j^{\mu}(x) = 0\,,~~j^{\mu} = \bar{\psi} \gamma^{\mu} \psi\,, 
\ee
one finds that the charge $Q$ is independent of time, where
\be\label{7}
 Q =\int dx\, j^{0}(x) =\int dx\, \psi^{\dag} \psi\,.
\ee
Further, the stress-energy tensor is also conserved:
\be\label{8}
 \partial_{\mu} T^{\mu\nu}(x) = 0\,,~~
 T_{\mu\nu} = \frac{1}{2}[D_{\mu\nu} + {h.c.}] - g_{\mu\nu} L\,,~~ 
D_{\mu\nu} = \bar{\psi} \gamma_{\mu} \partial_{\nu} \psi \>,   
\ee
leading to the result that linear momentum vector $P^{\nu} = (E,{P^{i}})$ is  time independent, where 
\be\label{9}
P_{\nu} = \int dx\,  T_{0 \nu} \,.
\ee
For stationary solutions  this leads to 
\be\label{10}
T_{10}= constant,~~ T_{11}= constant\,. 
\ee
We can write
\be\label{11}
T_{11}= \omega \psi^\dag \psi - m \bar{\psi} \psi + L_I\,. 
\ee
For solitary wave solutions vanishing at infinity the constant is zero so that
\be\label{12} 
\omega \psi^\dag \psi - m \bar{\psi} \psi + L_I =0\,.
\ee
On multiplying the equation of motion, Eq. (\ref{3a}), on the left by $\bar{\psi}$ we have 
\be\label{13}
2  L_I = -\omega \psi^{\dag} \psi + m \bar{\psi} \psi 
+ i \bar{\psi} \gamma_1 \partial_1 \psi\,.
\ee
Thus, 
\be\label{14}
\omega \psi^{\dag} \psi - m  \bar{\psi} \psi 
+ i\bar{\psi} \gamma_1 \partial_1 \psi =0\,.
\ee
We also have 
\be\label{15}
 L_I =   i \bar{\psi} \gamma_1 \partial_1 \psi \,, 
\ee
\be\label{16}
\bar{\psi} i\gamma_1 \partial_1 \psi = \psi^{\dag} \psi \frac{d\theta}{dx}\,.
\ee
This leads to the simple differential equation for $\theta$
\be\label{17}
\frac{d\theta}{dx} =  \kappa [m\cos(2\theta) -\omega]\,,
\ee
whose solution is 
\be \label{18}
\tan(\theta) = \alpha \tanh(\kappa \beta x)\,.
\ee
Here
\be\label{19}
\beta = \sqrt{m^2 - \omega^2}\,,~~\alpha = \sqrt{\frac{m-\omega}{m+\omega}}\,.
\ee
For $\kappa=1$  we have 
\be\label{21}
L_I = \frac{g^2}{2}R^{2  } \left[\frac{1}{q}\cos^{2}(2 \theta)
+\frac{1}{r}\sin^{2}(2\theta)+1/p\right]\,,
\ee
so that using Eq. (\ref{12}) we obtain
\be\label{22}
R^2[\omega-m\cos(2\theta)]+\frac{g^2 R^{4}}{2}
\left[\frac{1}{q}\cos^{2}(2\theta) +\frac{1}{r}\sin^{2 }(2\theta)+1/p\right] = 0\,.
\ee
This leads to
\be\label{23} 
R^2= \left[ \frac{2 (m \cos 2 \theta - \omega)}
{ g^2 \big (\frac{1}{q}\cos(2\theta)^{2} 
+ \frac{1}{r} \sin(2\theta)^{2}+1/p \big )} \right] .
\ee
As expected, in the limit $p,r \rightarrow \infty$, $R^2$ reduces to that of the
 SS case \cite{fred} while in the limit $r (p) \rightarrow \infty$ it reduces to
 generalized SS+VV or generalized SS+PP models, respectively. On using 
 Eq. (\ref{18})  and the identities
\be\label{24}
m+\omega\cosh(2 \beta x) = \frac{(m+\omega)
[1-\alpha^2 \tanh^2( \beta x)]}{\sech^2( \beta x)}\,,
\ee
\be\label{25}
\omega+m\cosh(2 \ \beta x) = \frac{(m+\omega)
[1+\alpha^2 \tanh^2( \ \beta x)]}{\sech^2(  \beta x)}\,,
\ee
one obtains an alternative expression for $R^2$ 
\be \label{rsqk1}
R^2= (1+\alpha^2 y^2) \bigg[\frac{2 (1-y^2) (m-\omega)}{g^2 
\big(\frac{1}{q}(1- \alpha^2 y^2)^2+\frac{1}{p}(1+\alpha^2 y^2)^2+
\frac{1}{r}(2\alpha y)^2\big)} \bigg] \,,  
\ee
where $y= \tanh( \beta x)$.

\section{Connection Between All Possible Combinations of SS, VV and 
PP NLD Interactions at $\kappa = 1$ With Various Combinations of  SS and VV}

We now show that at the special value of $\kappa = 1$, all possible combinations
of SS, VV and PP NLD interactions are in fact related to the various combinations 
of the NLD SS and VV interactions. The key point to note here is that
in $R^2$ as given by Eq. (\ref{rsqk1}) one has in the denominator different 
combinations of $(1+\alpha^2 y^2)$ or $(1-\alpha^2 y^2)$ or $4\alpha^2 y^2$ 
depending on whether one has VV or SS or PP interaction respectively where 
$y = \tanh(\beta x)$. Now observe that
\be\label{8.1}
4\alpha^2 y^2 = (1+\alpha^2 y^2)^2 - (1-\alpha^2 y^2)^2\,,
\ee
so that the PP interaction term can be converted to a combination of VV and 
SS interaction terms. This connection in $1+1$ dimensions could also have been 
found using the properties of the $\sigma$ matrices.  
We have, in general, that the denominator of  $R^2$ can be rewritten as
\bea 
&&g^2 
\left[\frac{1}{q}(1- \alpha^2 y^2)^2+\frac{1}{p} (1+\alpha^2 y^2)^2+
\frac{1}{r}(2\alpha y)^2\right]  \nonumber \\
&&=g^2 \left[(\frac{r-q}{qr})(1- \alpha^2 y^2)^2+(\frac{p+r}{pr})(1+\alpha^2 y^2)^2\right].
\eea
For the case $r=q$ we get a pure VV model and when $p+r=0$ we get the pure SS 
model. Otherwise, one can define 
\be\label{def} 
\tilde{q} = \frac{qr}{r-q} , ~~ \tilde{p} = \frac{pr} {p+r} \,, 
\ee
 to obtain the most general  $\pm VV \pm SS$
interaction, with 
\be \label{rsqk1b}
R^2= (1+\alpha^2 y^2) \bigg[\frac{2 (1-y^2) (m-\omega)}{g^2 
[\frac{1}{\tilde{q}}(1- \alpha^2 y^2)^2+\frac{1}{\tilde{p}}(1+\alpha^2 y^2)^2]}
\bigg] \,,  
\ee
where $y= \tanh( \beta x) $.  This corresponds to the class of interaction Lagrangians:

\be\label{2a}
 L_I = \frac{g^2}{2\tilde{q}}(\bar{\psi} \psi)^{2}+\frac{g^2}{2 \tilde{p}}[\bar{\psi} \gamma_{\mu}\psi 
\bar{\psi} \gamma^{\mu} \psi]. 
\ee

\subsection{Charge and Energy of the Solitary Waves} \label{QE} 

It is straightforward to compute the charge $Q$ using Eqs. (\ref{7}) and 
(\ref{rsqk1}). On making the substitution $y = \tanh(\beta x)$ we obtain
\be\label{Q}
Q = \frac{4  \alpha}{g^2}   I(\omega,,p,q,r )\,,
\ee
where
\be\label{I}
I(\omega,p,q,r) = \int_{0}^{1} dy~  \frac{(1+ \alpha^2 y^2)}
{[\frac{1}{q}(1- \alpha^2 y^2)^{2}+\frac{1}{p}
(1+\alpha^2 y^2)^{2}+\frac{1}{r}(2 \alpha y)^{2}]}\,.
\ee

For the energy we have that the Hamiltonian density is given by 
\be\label{3.12}
h = T_{00} =  i \bar{\psi} \gamma_1 \partial_1 \psi+ m \bar{\psi} \psi -L_I
\equiv  h_1+h_2-h_3\,.
\ee
From Eq. (\ref{15}) it follows that $h_3=h_1$ , and hence
\be\label{3.13}
h=h_2\,.
\ee
Thus the total energy $E$ is given by
\be\label{3.16}
E= H_2 = \int_{-\infty}^{\infty} dx\, R^2(x) \cos(2\theta)\,. 
\ee
This can be written as:
\bea\label{20}
&&E = \int_{-\infty}^{\infty} dx R^2(x) \cos(2\theta) \nonumber \\
&& = \frac{4m\alpha}{g^2} \int_{0}^{1} dy\, \frac{(1-\alpha^2 y^2)}
{[\frac{(1-\alpha^2 y^2)^{2}}{q} +\frac{(1+\alpha^2 y^2)^{2}}{p} 
+\frac{4 \alpha^2 y^2}{r}]}\,. 
\eea
We see that $E/(mQ)$ is independent of $g^2$ but depends on $p,q,r$.  When 
$E/(mQ) <1$, which will be the case at $\kappa=1$, then all the solutions are 
bound states. One needs to ensure the denominator does not blow up in the 
range $0 \leq y \leq 1$.

At $\kappa=1$ we can directly evaluate $E$  and $Q$ in terms of simple functions.  So we can explicitly  show that at fixed 
$p,q,r$ we have  (setting $m=1$) 
\be \label{dedw}
\frac{dE(\omega) } {d \omega} =  \omega  \frac{dQ}{d\omega}.
\ee
Also one finds that  
\be \label{ineq}
\frac{E(\omega) }{Q (\omega) } > \omega.
\ee
Thus one finds  the slope of $E/Q$ wlll have the opposite sign to the slope of $Q(\omega)$ since
\bea
 \frac{d}{d \omega} \left( \frac{E}{Q}\right) &&= \frac{E'(\omega)} {Q(\omega)} - E(\omega) \frac{Q'(\omega)} {Q(\omega)}  \nonumber \\
&&=\left(\omega-\frac{E}{Q} \right)  \frac{Q'}{Q} \,. 
\eea
(Here the prime signifies $d/d \omega$). 
Since $Q >0$ this shows using Eqs. (\ref{dedw},\ref{ineq}) that the slope of $E/Q$ is opposite of the slope of
$dQ/d\omega$, which is the slope related to the Vakhitov-Kolokolov stability for the NLSE.  So, positivity of this slope
(which is always the case at $\kappa=1$) guarantees linear stability in the non-relativistic regime (see also \cite{sqm} and the Appendix).

\subsubsection{Possible restrictions on $\omega$ in order that the charge $Q$ exists}

For simplicity, throughout this paper we shall be measuring $\omega$ and $E$ in units of $m$. 
For the integral in Eq. (\ref{I}) to converge at $0 \le y \le 1$, the 
denominator must not vanish when $0 \le y \le 1$.  For $p, q, r > 0$  this is 
always true. The general condition is easily shown to be
\be
\frac{4 \left(p \left(-q\omega^2+q+r \omega^2\right)+q r\right)}{p q r (\omega+1)^2} >0.
 \ee
Using this formula we can find the  value of $\omega$  ($\omega_{min}$) 
 above which  a solitary wave will  exist. 
There will be a minimum  value of $\omega$ if $\omega <1$ and we can satisfy 
 \be \label{omin}
0 < \omega_{min} = \sqrt{\frac{q(p+r)}{p(q-r)}}  <1 .
\ee  
 When $\omega_{min}$ is real and is in the allowed regime then it implies that
the solitary waves will exist only when 
 \be    
0<  \omega_{min} < \omega \leq 1\,.
 \ee
It is easy to check that Eq. (\ref{omin}) correctly reproduces the well-known
results on $\omega_{min}$ in all the five NLD models with admixtures of of SS 
and VV interactions.

\subsection{Double-Hump vs Single-Hump Behavior of $R^2$}

Depending on the values of $p,q,r$ the solitary wave will have either one or 
two humps.  At $x=0$ the first derivative $d(R^2)/dx =0$. 
There will be one or two humps if $R^2$ has a maximum or minimum at $x=0$, 
i.e whether  the second derivative  of $R^2$ with respect to $x$ is negative or
positive at $x=0$. As a function of $\omega$ when there are double hump 
solitary wave solutions, there will be a critical value of $\omega$ when the
derivative is zero which will signify the transition from the double-hump to 
the single-hump solitary waves as a function of $\omega$. At that point the 
solitary wave is flat topped. We can rewrite $R^2$ in the form: 
\be\label{3.4a}
R^2/C = (1+\alpha^2 T^2)\frac{S^2}
{[\frac{1}{q}(1-\alpha^2 T^2)^{2} +\frac{1}{p}(1+\alpha^2 T^2)^{2} 
+ \frac{1}{r}(2\alpha T)^{2}]}\,,
\ee
where $T$ and $S$ stand for $\tanh(\beta x)$ and $\sech(\beta x)$, respectively, 
while in $C$ we have absorbed all inessential $x$ independent constants.

We find the first derivative is zero at $x=0$ and the 
second derivative at $x=0$  (apart from a constant) is given by
\bq
 \frac{d^2 R^2}{dx^2}\Big |_{x=0} = -\frac{4 p q (\omega -1) (p (2 q (\omega -1)-2 r
 \omega +r)-q r)}{r (p+q)^2}\,.
\eq
This is zero when
\bq\label{dh}
\omega =\omega_0 =  \frac{2 p q-p r+q r}{2 p (q-r)}.
\eq
So, we see if $\omega_0$  is real and in the range $0 < \omega_0 <1$, one can 
have double-humped solutions for $\omega_{min} \le \omega \le \omega_0$. On the
other hand, when $\omega_0  < \omega <1$ the solitary waves have a single hump.
It is easy to check that Eq. (\ref{dh}) correctly reproduces the well-known
results on $\omega_{0}$ in all the five NLD models with admixtures of of SS 
and VV interactions.

\section{All Possible Combinations of SS, VV and PP Interactions at $\kappa = 1$} 

In previous papers we have worked out the various features of the solitary wave
solutions  of the $SS$, $VV$ \cite{fred}, SS$-$VV \cite{ak}, the VV$-$SS  
\cite{cgabs} and the SS$+$VV models \cite{fred7}. In this section we will relate 
the various possible choices of SS-VV-PP Interactions at $\kappa = 1$ to the
well-known combinations of SS and VV Interactions. In each case, we mention
(i) the restriction on $\omega$, (ii) existence of double hump and single hump
solitary wave solutions, and (iii) explicit expression for the charge $Q$ and energy $E$.

\subsection{Case I: SS$-$VV$-$PP Interaction}

Here we take $q=1 , p < -1, r < 0$ so that at $\kappa = 1$, the denominator of  
$R^2$ as given by Eq. (\ref{rsqk1}) takes the form
\be\label{1.2}
\frac{1}{g^2}\left[(1-\alpha^2 y^2)^2 -\frac{(1+\alpha^2 y^2)^2}{|p|} 
-\frac{4\alpha^2 y^2}{|r|}\right]\,,~~ |p| > 1, |r| > 0\,,
\ee
which  can be re-expressed as
\be\label{1.3}
\frac{1}{(\hat{g})^2}\left[(1-\alpha^2 y^2)^2 -\frac{(1+\alpha^2 y^2)^2}
{|\hat{p}|}\right]
\equiv  \frac{1}{(\hat{g})^2}[SS-\frac{1}{\hat{p}} VV]\,.
\ee
Here
\be\label{1.4}
(\hat{g})^2 = \frac{|r|g^2}{(|r|+1)} < g^2\,,~~\hat{p} = \frac{|p|(|r|+1)}
{(|p|+|r|)} > 1\,.
\ee
This is in the form of the $SS-VV$ model that we have studied earlier 
\cite{ak}. Using our general formulas for  $\omega_{min}$ or the results of 
\cite{ak}  we find that
\bq
\sqrt{\frac{(|p|+|r|)}{|p|(|r|+1)}} \le \omega < 1\,.
\eq
Further, for the Case I, one predicts that there will be solitary wave 
solutions with a double hump in case
\be\label{1.6}
\sqrt{\frac{(|p|+|r|)}{|p|(|r|+1)}} \le \omega \le
\frac{|p|(|r|+2)+|r|}{2|p|(|r|+1)}\,,
\ee
while there will be solitary wave solutions with a single hump in case
\be\label{1.7}
\frac{|p|(|r|+2)+|r|}{2|p|(|r|+1)} < \omega < 1\,.
\ee

\begin{figure} 
    \centering
    \includegraphics[width=0.7 \linewidth]{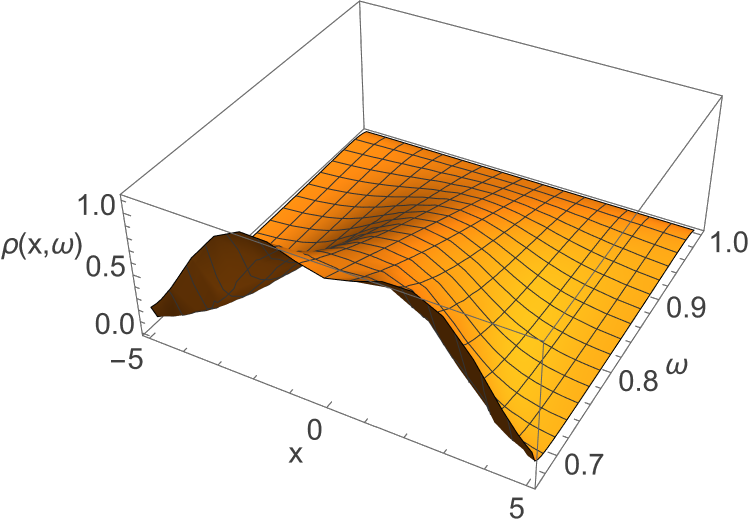}
   \caption{ \label{rsqmm35} Plot of $R^2(x,\omega) $ when $\kappa=1,q=1, p=-3,
   r=-5$. }
\end{figure}

For  $p=-3, r=-5$ we find that $\omega_{min} = 2/3$.  For  this value we
display $R^2(\omega)$ in Fig. (\ref{rsqmm35}).  The transition from two humps 
to one hump takes place at $\omega_0= \frac{13}{18} \approx 0.7222$.

\begin{figure}[t]
    \centering
    \includegraphics[width=0.7 \linewidth]{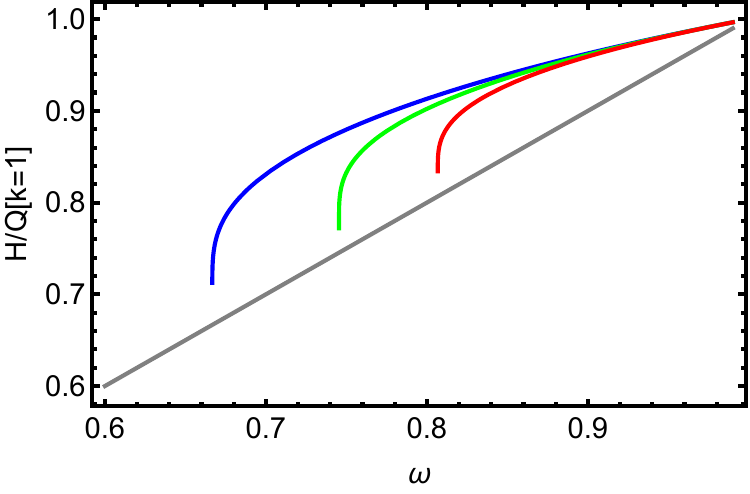}
    \caption{\label{EQk1til}  Plot of $E(\omega,p,r)/Q(\omega,p,r)$ when  $\kappa=1$ for   $p= - 3$ and $r=-1.1$ (red), $-2$ (green) , $-5$ (blue) compared with 
    $\omega$ (gray).}    %
\end{figure}

Finally, using Eqs. (\ref{18}) and (\ref{20}) we find that for the Case I
\be\label{1.8}
Q = \frac{2|p|(|r|+1)^2}{g^2|r|\sqrt{|r|(|p|-1)(|p|+|r|)}}
\tanh^{-1}\bigg [\frac{\beta\sqrt{|p|+|r|}}{\omega\sqrt{|r|(|p|-1)}} \bigg]\,,
\ee
\be\label{1.9}
E = \frac{2(|r|+1)\sqrt{|p|(|r|+1)}}{g^2|r|\sqrt{|r|(|p|-1)}}
\tanh^{-1}\bigg [\frac{\beta\sqrt{|p|(|r|+1)}}{\sqrt{|r|(|p|-1)}}\bigg]\,.
\ee
For the case $p=-3$ and $r<0$ as we increase $|r|$ then the range of $\omega$ 
increases. A plot of $E/Q$ as a function of  $\omega$ for $\kappa=1, p=-3, 
r= -1.2,-3- 5 $  is shown in Fig.~\ref{EQk1til} where we see this phenomenon. 
Since $E/Q <  1$,  all these solutions are bound states.  We see that as discussed earlier $E/Q>\omega$.   We expect that  these solitary waves are stable in the non-relativistic regime  because the slope of $E/Q$  is always positive.  
Note the generic behavior when there is an $\omega_{0} > \omega_{min}$ such that $E/Q$ has an inflection point  at  $\omega_{0}$ where the slope changes as we change from two-hump to one-hump solitary waves.  

\subsection{Case II: SS$-$PP Interaction}

In the limit of $|p| \rightarrow \infty$, case I reduces to the $SS-PP$ model 
which can  be
re-expressed as $SS-VV$ model with  $\hat{p} \rightarrow (|r|+1)$. 
Thus, in this case $\omega$ is restricted to
\be\label{1.10}
\frac{1}{\sqrt{|r|+1}} \le \omega < 1\,.
\ee
Further, there are solitary wave solutions with a double hump in case
\be\label{1.11}
\frac{1}{\sqrt{|r|+1}} \le \omega \le \frac{(|r|+2)}{2(|r|+1)}\,,
\ee
while there are solitary wave solutions with a single hump in case
\be\label{1.12}
\frac{(|r|+2)}{2(|r|+1)} < \omega < 1\,.
\ee
Finally, in this case
\be\label{1.13}
Q = \frac{2(|r|+1)^2}{g^2|r|\sqrt{|r|}}
\tanh^{-1}\left[\frac{\beta}{\omega\sqrt{|r|}}\right]\,,
\ee
\be\label{1.14}
E = \frac{2(|r|+1)\sqrt{|r|+1}}{g^2|r|\sqrt{|r|}}
\tanh^{-1}\bigg [\frac{\beta}{\sqrt{|r|}} \bigg]\,.
\ee
Here $\alpha, \beta$ are as given by Eq. (\ref{12}) with $m = 1$.

\subsection{ Case III: VV$-$SS$+$PP Interaction}
\begin{figure}  
    \centering
    \includegraphics[width=0.7 \linewidth]{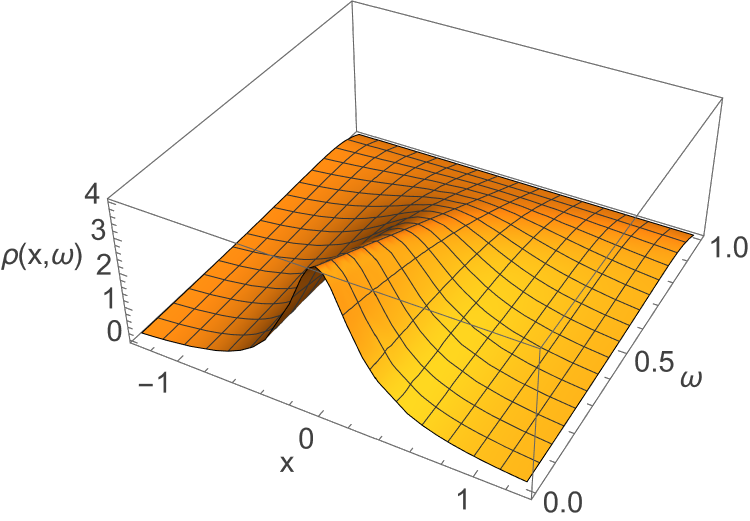}]
    \caption{ \label{rhop1qm2r1}  Plot of   $R^2(\omega,x) $    when  $p=1,q=-2,r=1$. }
   \end{figure}

In this case $p = 1, q < -1, r > 0$ so that at $\kappa = 1, R^2$ has a term
\be\label{1.15}
g^2\left[(1+\alpha^2 y^2)^2 -\frac{(1-\alpha^2 y^2)^2}{|q|} 
+\frac{4\alpha^2 y^2}{r}\right]\,,~~ |q| > 1, r > 0\,,
\ee
which  can be re-expressed as
\be\label{1.16}
\frac{1}{(\hat{g})^2}\left[(1+\alpha^2 y^2)^2 -\frac{(1-\alpha^2 y^2)^2}
{\hat{q}}\right] 
 \equiv \frac{1}{(\hat{g})^2}\left[VV-\frac{1}{\hat{q}} SS\right]\,.
\ee
where
\be\label{1.17}
(\hat{g})^2 = \frac{r g^2}{(r+1)} < g^2\,,~~\hat{q} = \frac{|q|(r+1)}
{(|q|+r)} > 1\,.
\ee
Using the results for the $VV-SS$ model \cite{cgabs} we find
 that in this case there is no restriction on $\omega$,
$0 < \omega < 1$. Furthermore, in this case all the solitary waves have only a
single hump.
An example of this case is shown in Fig. (\ref{rhop1qm2r1}) where we have 
chosen $p=1,q=-2$ and $ r=1$. 

Finally, using Eqs. (\ref{18}) and (\ref{20}) we find that for the Case III
\be\label{1.18}
Q = \frac{2(r+1)\sqrt{|q|(r+1)}} {g^2 r\sqrt{r(|q|-1)}}
\tan^{-1}\bigg [\frac{\beta\sqrt{|q|(1+r)}} {\omega\sqrt{r(|q|-1)}} \bigg]\,,
\ee
\be\label{1.19}
E = \frac{2 |q|(r+1)^2}{g^2 r\sqrt{r(|q|-1)(|q|+r)}}
\tan^{-1}\bigg [\frac{\beta\sqrt{(|q|+r)}}{\sqrt{r(|q|-1)}}\bigg]\,.
\ee

\subsection{Case IV: VV$+$PP Interaction}

In the limit $q \rightarrow \infty$, Case III reduces to that of VV$+$PP.
Thus, it follows that VV$+$PP model can also be re-expressed as VV$-$SS model 
with $\hat{q} =  - r <0$. Therefore, in this case too there is no restriction on 
$\omega$, $0 < \omega < 1$. Furthermore, in this case all the solitary waves 
have only a single hump. Finally, in this case
\be\label{1.20}
Q = \frac{2(r+1)\sqrt{(r+1)}}{g^2 r\sqrt{r}}
\tan^{-1}\bigg [\frac{\beta\sqrt{(1+r)}}{\omega\sqrt{r}} \bigg]\,,
\ee
\be\label{1.21}
E = \frac{2(r+1)^2}{g^2 r\sqrt{r}}
\tan^{-1}\bigg [\frac{\beta}{\sqrt{r}}\bigg]\,.
\ee

\subsection{Case V: SS$-$VV$+$PP Interaction}  

In this case $q = 1, p < -1$ and $r > 0$ and at $\kappa = 1$, $R^2$ has a term 
in the denominator of the form
\be\label{1.22}
g^2 \left[(1-\alpha^2 y^2)^2 -\frac{(1+\alpha^2 y^2)^2}{|p|} 
+\frac{4\alpha^2 y^2}{r}\right]\,,~~ |p| > 1, r > 0\,,
\ee
which can be re-expressed as
\be\label{1.23}
g^2 \left[\frac{(r-1)}{r}(1-\alpha^2 y^2)^2 +\frac{(|p|-r)}{|p|r} 
(1+\alpha^2 y^2)^2\right]\,.
\ee
Now depending on the values of the parameters $r > 0, |p| > 1$ one gets
different combinations of SS and VV NLD interactions which we discuss one by 
one. 

\subsubsection{\bf $r = 1$}

In this case one finds that
\be\label{1.24}
SS-VV+PP = \frac{1}{(\hat{g})^2} VV\,,~~(\hat{g})^2 = \frac{|p| g^2}{(|p|-1)} 
> g^2\,.
\ee
Using the results for the $VV$ model \cite{fred} one finds that  there is no 
restriction on $\omega$ and $0 < \omega < 1$. For this case all the solitary 
waves have only a single hump. Finally, for this case 
\be\label{1.25}
Q = \frac{4(|p|-1)}{|p| g^2} \tan^{-1}(\alpha)\,,~~
E = \frac{2(|p|-1)\beta}{|p|g^2}\,.
\ee

\subsubsection{\bf $0 < r < 1$}

In this case one finds
\bea\label{1.26}
&&SS-VV+PP = \frac{1}{(\hat{g})^2} [VV -\frac{1}{\hat{q}} SS]\,, 
\nonumber \\
&&(\hat{g})^2 = \frac{|p|r g^2}{(|p|-r)}\,, ~~\hat{q} 
= \frac{(|p|-r)}{|p|(1-r)} > 1\,. 
\eea
Using the  results for the VV$-$SS model \cite{cgabs} we then 
predict that in this case there is no restriction on $\omega$ and 
$0 < \omega < 1$. Futher, in this case all the solitary waves have only 
a single hump. Finally, for this case
\be\label{1.28}
Q = \frac{2(|p|-r)\sqrt{(|p|-r)}}{g^2 |p| r\sqrt{r(|p|-1)}}
\tan^{-1}\left(\frac{\beta\sqrt{|p|-r}}{\omega \sqrt{r(|p|-1)}} \right )\,.
\ee
\be\label{1.29}
E = \frac{2(|p|-r)^2}{g^2 |p| r \sqrt{|p| r (|p|-1)(1-r)}}
\tan^{-1}\left(\frac{\beta \sqrt{|p|(1-r)}}{\sqrt{r(|p|-1)}} \right )\,.
\ee

 \subsubsection{\bf $r = |p| > 1$}

In this case
\be\label{1.30}
SS-VV+PP = \frac{1}{(\hat{g})^2} SS\,, ~~(\hat{g})^2 = \frac{r g^2}{(r-1)} > 
g^2\,. 
\ee
Using the results of the $SS$ model \cite{fred} we find that there is no 
restriction on $\omega$ and $0 < \omega < 1$. However,
all the solitary waves in this case have a double hump in case 
$0 < \omega \le 1/2$ and a single hump when  $1/2  < \omega <1$. 
Finally, in this case
\be\label{1.33}
Q = \frac{2(r-1)\beta}{r g^2 \omega}\,.
\ee
\be\label{1.34}
E = \frac{2(r-1)}{g^2 r} \tanh^{-1}(\beta)\,.
\ee

\subsubsection{\bf $r > |p| > 1$}

In this case
\bea\label{1.35}
&&SS-VV+PP = \frac{1}{(\hat{g})^2} [SS -\frac{1}{\hat{p}} VV]\,, 
\nonumber \\
&&(\hat{g})^2 = \frac{r g^2}{(r-1)} > g^2\,, ~~\hat{p} 
= \frac{|p|(r-1)}{(r -|p|)} > 1\,. 
\eea
Using the results of the $SS-VV$ model \cite{ak} we find $\omega$ is restricted to
\be\label{1.36}
\sqrt{\frac{(r-|p|)}{|p|(r-1)}} \le \omega < 1\,.
\ee
The solitary waves in this case have two humps in case
\be\label{1.37}
\sqrt{\frac{(r-|p|)}{|p|(r-1)}} \le \omega < \frac{r(|p|-1)}
{2|p|(r-1)}\,,
\ee
while they have a single hump in case
\be\label{1.38}
\frac{r(|p|-1)}{2|p|(r-1)} < \omega < 1\,.
\ee

For $p=-3, r=5$ one gets the transition from a double hump to a single hump 
solitary waves when $\omega_c = \frac{7}{12}$. The behavior of $R^2(\omega)$ 
for this value of $\omega_c$  is shown in Fig. (\ref{rsqpp35}) where we see the
generic behavior that at $\omega_c$ the solitary wave has a flat top. 
\begin{figure}  
    \centering
    \includegraphics[width=0.7 \linewidth]{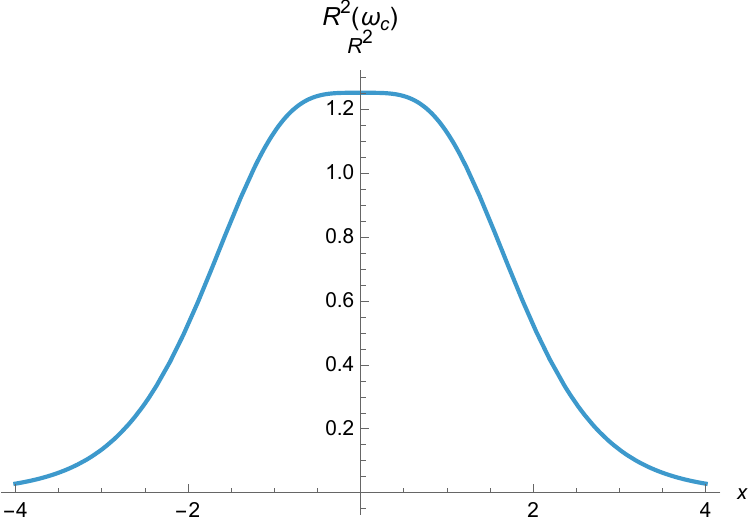}
    \caption{ \label{rsqpp35}  Plot of   $R^2(\omega) $    when  $\kappa=1, p=-3,r=5$, $\omega_c=7/12$. }
\end{figure}

\begin{figure} 
    \centering
    \includegraphics[width=0.7 \linewidth]{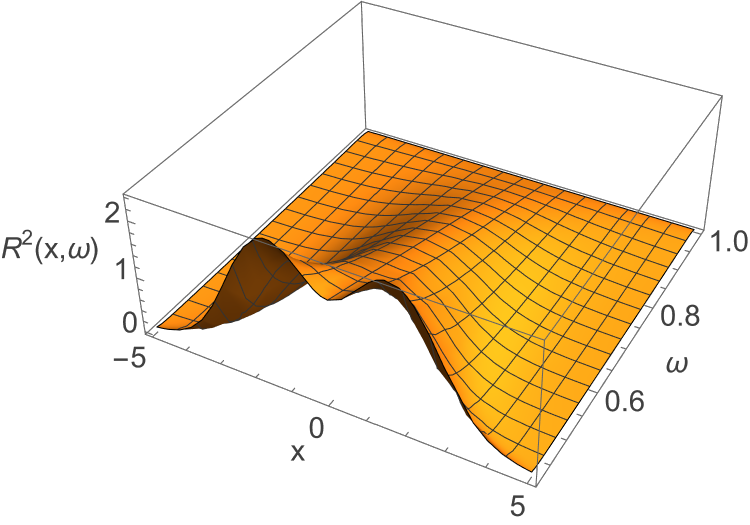}
    \caption{ \label{rsqabs35}  Plot of   $R^2(x,\omega) $    when  $\kappa=1, p=-3,r=5$. }
\end{figure}

There is also a minimum value of $\omega$ for this choice:
\be
\omega_{min} = \frac{1}{\sqrt{6}} \,. 
\ee
We can see the transition from double to single hump solitary wave in 
Fig. (\ref{rsqabs35})

Finally, in this case
\be\label{1.39}
Q = \frac{2|p|(r-1)\sqrt{(r-1)}}{r g^2\sqrt{r (|p|-1)(r-|p|)}}
\tanh^{-1} \bigg[\frac{\beta\sqrt{(r-|p|)}}{\omega \sqrt{r(|p|-1)}} \bigg ]\,,
\ee
\be\label{1.40}
E = \frac{2(r-1)\sqrt{(r-1)|p|}}{r g^2\sqrt{r(|p|-1)}}
\tanh^{-1} \bigg[\frac{\beta\sqrt{|p|(r-1)}}{\sqrt{r(|p|-1)}} \bigg]\,.
\ee

\subsubsection{\bf $|p| > r > 1$}

In this case
\bea\label{1.41}
&&SS-VV+PP = \frac{1}{(\hat{g})^2} [SS +\frac{1}{\hat{p}} VV]\,, 
\nonumber \\
&&(\hat{g})^2 = \frac{r g^2}{(r-1)} > g^2\,, ~~\hat{p} 
= \frac{|p|(r-1)}{(|p|-r)} > 0\,. 
\eea
Using the  results of the $SS+VV$ model \cite{fred7} we then predict
that in this case there is no restriction on $\omega$ and $0 < \omega < m$ is
allowed. Further, all the solitary waves in this case have a double hump in case
\be\label{1.42}
0 < \frac{\omega}{m} \le \frac{|p|(r-2)+r}{2|p|(r-1)}\,,
\ee
while they have a single hump in case
\be\label{1.43}
\frac{|p|(r-2)+ r}{2|p|(r-1)} < \omega < 1\,.
\ee
Finally, in this case
\be\label{1.44}
Q = \frac{2|p|(r-1)^2}{g^2 r\sqrt{r(|p|-1)(|p|-r)}}
\tan^{-1} \bigg[\frac{\beta\sqrt{(|p|-r)}}{\omega \sqrt{r(|p|-1)}} \bigg]\,,
\ee
\be\label{1.45}
E = \frac{2(r-1)\sqrt{(r-1)|p|}}{g^2 r \sqrt{r(|p|-1)}}
\tanh^{-1} \bigg[\frac{\beta\sqrt{|p|(r-1)}}{\sqrt{r(|p|-1)}} \bigg]\,.
\ee

\subsection{\bf Solution VI: SS+PP Model}

In the limit of $|p| \rightarrow \infty$, the Case V reduces to that of the
SS+PP model. Thus, it follows that at $\kappa = 1$, $SS+PP$ case can also be
re-expressed as $SS+VV$ model. In particular,  as $p \rightarrow \infty$, the 
$SS+PP$ model at $\kappa = 1$ can be re-expressed as
\be\label{1.46}
SS+PP= \frac{1}{g^2}\left[\frac{(r-1)}{r}(1-\alpha^2 y^2)^2 +\frac{1}{r} 
(1+\alpha^2 y^2)^2\right]\,.
\ee
Now depending on the values of the parameters $r > 0$ one gets
different combinations of SS and VV NLD interactions. We obtain

\subsubsection{\bf $r = 1$}

In this case
\be\label{1.47}
SS+PP = \frac{1}{g^2} VV\,.
\ee
From the results for the  $VV$ model \cite{fred} we conclude that there is no 
restriction on $\omega$ and $0 < \omega < 1$. Furthermore, in this case all 
the solitary waves have only a single hump. Finally, in this case 
\be\label{1.48}
Q = \frac{4 \tan^{-1}(\alpha)}{g^2} , ~~
E =  \frac{2\beta}{g^2}\,.
\ee

\subsubsection{\bf $0 < r < 1$}

In this case
\bea\label{1.49}
&&SS+PP = \frac{1}{(\hat{g})^2} [VV -\frac{1}{\hat{q}} SS]\,, 
\nonumber \\
&&(\hat{g})^2 = r g^2\,, ~~\hat{q} = \frac{1}{(1-r)} > 1\,. 
\eea
Using the results  for the $VV-SS$ model \cite{cgabs} we find there is no 
restriction on $\omega$ and $0 < \omega < 1$. For this case all the solitary 
waves have only a single hump. Finally, for this case
\be\label{1.50}
Q = \frac{2}{r \sqrt{r} g^2}
\tan^{-1}\left(\frac{\beta}{\omega \sqrt{r}} \right )\,,
\ee
\be\label{1.51}
E = \frac{2}{r g^2\sqrt{r(1-r)}}
\tan^{-1}\left(\frac{\beta}{\sqrt{r}} \right )\,.
\ee

\subsubsection{\bf $r > 1$}

In this case
\bea\label{1.52}
&&SS+PP = \frac{1}{(\hat{g})^2} [SS +\frac{1}{\hat{p}} VV]\,, 
\nonumber \\
&&(\hat{g})^2 = \frac{r g^2}{(r-1)} > g^2\,, ~~\hat{p} 
= (r-1) > 0\,. 
\eea
Using the  results of the $SS+VV$ model \cite{fred7} we find
that in this case there is no restriction on $\omega$ and $0 < \omega < 1$ is
allowed. Further, all the solitary waves in this case have two humps in case
\be\label{1.53}
0 < \omega \le \frac{(r-2)}{2(r-1)}\,,
\ee
while they have a single hump in case
\be\label{1.54}
\frac{(r-2)}{2(r-1)} < \omega < 1\,.
\ee
Thus in case $1 < r \le 2$, then all solitary waves have only a single hump.
Finally, in this case
\be\label{1.55}
Q = \frac{2(r-1)^2}{g^2 r\sqrt{r}}
\tan^{-1} \left(\frac{\beta}{\omega \sqrt{r}} \right)\, 
\ee
\be\label{1.56}
E = \frac{2(r-1)\sqrt{(r-1)}}{g^2 r \sqrt{r}}
\tanh^{-1} \left(\frac{\beta}{\sqrt{r}} \right)\,.
\ee

\subsection{Case VII: SS+VV$-$PP Interaction} 

In this case at $\kappa = 1$, $q=1$,  $p > 0, r < 0$  and 
 $R^2$ has a term in the denominator
\be\label{1.57}
g^2\left[(1-\alpha^2 y^2)^2 +\frac{(1+\alpha^2 y^2)^2}{p} 
-\frac{4\alpha^2 y^2}{|r|}\right]\,,~~ p, |r| > 0\,,
\ee
which can be reexpressed as
\be\label{1.58}
g^2 [ \frac{(|r|+1)}{|r|}(1-\alpha^2 y^2)^2 +\frac{(|r|-p)}{|r|p} 
(1+\alpha^2 y^2)^2]\,.
\ee
Now depending on the values of the parameters $p, |r| > 0$ one gets
different combinations of SS and VV NLD interactions. 

\subsubsection{\bf $|r| = p$}

In this case
\be\label{1.59}
SS+VV-PP = \frac{1}{(\hat{g})^2} SS\,,~~(\hat{g})^2 = \frac{|r| g^2}{(|r|+1)} 
< g^2\,.
\ee
Using the  results of the SS model \cite{fred} we then predict
that in this case there is no restriction on $\omega$ and it can take any
value $0 < \omega < 1$.
Further, the solitary waves in this case have a double hump in case
\be\label{1.60}
0 < \omega < 1/2\,,
\ee
while they have a single hump in case
\be\label{1.61}
1/2 < \omega < 1\,.
\ee
In this case we obtain
\be\label{1.62}
Q = \frac{2(|r|+1)\beta}{|r| g^2 \omega}\,.
\ee
\be\label{1.63}
E = \frac{2(|r|+1)\beta}{|r| g^2}\,.
\ee

\subsubsection{\bf $p > |r|$}

In this case
\bea\label{1.64}
&&SS+VV-PP = \frac{1}{(\hat{g})^2} [SS -\frac{1}{\hat{p}} VV]\,, 
\nonumber \\
&&(\hat{g})^2 = \frac{|r| g^2}{(|r|+1)} < g^2\,, ~~\hat{p} 
= \frac{p(|r|+1)}{(p-|r|)} > 1\,. 
\eea
Using the  results of the $SS-VV$ model \cite{ak} we then predict
that in this case $\omega$ is restricted to
\be\label{1.65}
\sqrt{\frac{(p-|r|)}{p(|r|+1)}} < \omega < 1\,.
\ee
Further, all the solitary waves in this case have a double hump in case
\be\label{1.66}
\sqrt{\frac{(p-|r|)}{p(|r|+1)}} < \omega < \frac{p(|r|+2)-|r|}
{2p(|r|+1)}\,,
\ee
while they have a single hump in case
\be\label{1.67}
\frac{p(|r|+2)-|r|}{2p(|r|+1)} < \omega < 1\,.
\ee
In this case we obtain
\be\label{1.68}
Q = \frac{2p(|r|+1)^2}{|r| g^2\sqrt{|r| (p+1)(p-|r|)}}
\tanh^{-1}\left(\frac{\beta \sqrt{p-|r|}}{\omega \sqrt{|r|(p+1)}} \right )\,, 
\ee
\be\label{1.69}
E = \frac{2(|r|+1)\sqrt{p(|r|+1)}}{g^2 |r|\sqrt{(p+1)|r|}}
\tanh^{-1}\left(\frac{\beta \sqrt{p(|r|+1}}{\sqrt{|r|(p+1)}} \right )\,.
\ee

\subsubsection{\bf $|r| > p > 1$}

In this case
\bea\label{1.70}
&&SS+VV-PP = \frac{1}{(\hat{g})^2} [SS +\frac{1}{|\hat{p}|} VV]\,, 
\nonumber \\
&&(\hat{g})^2 = \frac{|r| g^2}{(|r|+1)} < g^2\,, ~~\hat{p} 
= \frac{p(|r|+1)}{(|r|-p)} > 0\,. 
\eea
Using the  results of the $SS+VV$ model \cite{fred7} we then predict
that in this case there is no restriction on $\omega$ and $0 < \omega < 1$ is
allowed. Further, all the solitary waves in this case have a double hump in case
\be\label{1.71}
0 <  \omega < \frac{p(|r|+2)-|r|}{2p(|r|+1)}\,,
\ee
while they have single hump in case
\be\label{1.72}
\frac{p(|r|+2)- |r|}{2p(|r|+1)} < \omega < 1\,.
\ee

\begin{figure} 
    \centering
    \includegraphics[width=0.7 \linewidth]{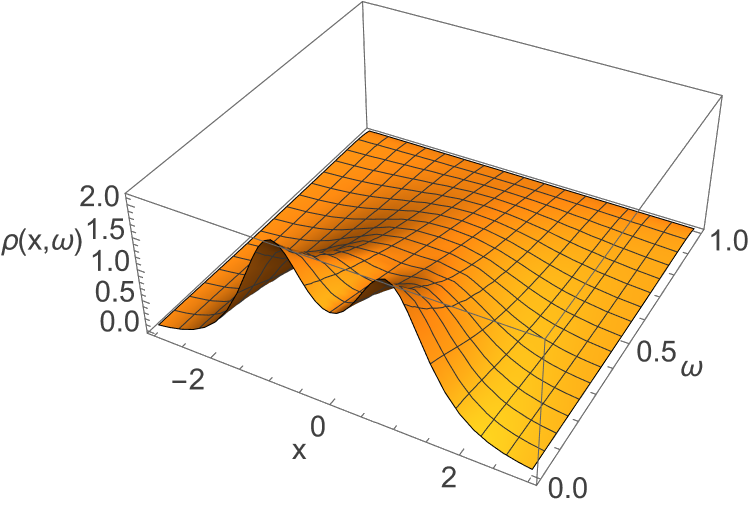}
    \caption{ \label{p2r-3}  Plot of   $R^2(x,\omega) $    when  $\kappa=1, p=2,r=-3$. }
\end{figure}

Choosing $p=2,r=-3$, we have $\omega_0= 7/16$ and we get the result for 
$R^2(x, \omega)$ shown in Fig. (\ref{p2r-3}).
In this case we obtain
\be\label{1.73}
Q = \frac{2p (|r|+1)^2}{g^2 |r|\sqrt{|r| (p+1)(|r|-p)}}
\tan^{-1}\left(\frac{\beta \sqrt{p(|r|-p)}}{\omega \sqrt{|r|(p+1)}} \right )\,,
\ee
\be\label{1.74}
E = \frac{2 (|r|+1)\sqrt{(|r|+1)}}{g^2 |r| \sqrt{(p+1)|r|}}
\tanh^{-1}\left(\frac{\beta \sqrt{p(|r|+1}}{\sqrt{|r|(p+1)}} \right )\,.
\ee

\subsection{Case VIII: SS+VV+PP Interactions}

In this case $q=1$,  $p, r > 0$ and at $\kappa = 1$, $R^2$ has the term
\be\label{1.75}
g^2 \left[(1-\alpha^2 y^2)^2 +\frac{(1+\alpha^2 y^2)^2}{p} 
+\frac{4\alpha^2 y^2}{r}\right]\,,~~ p, r > 0\,,
\ee
which can be reexpressed as
\be\label{1.76}
\frac{1}{g^2}\left[\frac{(r-1)}{r}(1-\alpha^2 y^2)^2 +\frac{(r+p)}{rp} 
(1+\alpha^2 y^2)^2\right]\,.
\ee
Now depending on the values of the parameters $p, r > 0$ one gets
different combinations of SS and VV NLD interactions. 

\subsubsection{\bf $r = 1$}

In this case
\be\label{1.77}
SS+VV+PP = \frac{1}{(\hat{g})^2} VV\,,~~(\hat{g})^2 = \frac{p g^2}{p+1} 
< g^2\,.
\ee
Using the  results of the $VV$ model \cite{fred} we then predict
that in this case there is no restriction on $\omega$ and $0 < \omega < 1$ is
allowed. Further, all the solitary waves in this case have single hump.
In this case we find
\be\label{1.78}
Q = \frac{4(p+1)}{g^2 p} \tan^{-1}(\alpha)\,,
\ee
\be\label{1.79}
E = \frac{2(p+1)\beta}{g^2 p}\,.
\ee

\subsubsection{\bf $0 < r < 1$}

In this case
\bea\label{1.80}
&&SS+VV+PP = \frac{1}{(\hat{g})^2} [VV -\frac{1}{|\hat{q}|} SS]\,, 
\nonumber \\
&&(\hat{g})^2 = \frac{p r g^2}{(p+r)} < g^2\,, ~~\hat{q} 
= \frac{(p+r)}{p(1-r)} > 1\,. 
\eea
Using the  results of the $VV-SS$ model \cite{cgabs} we then predict
that in this case there is no restriction on $\omega$ and $0 < \omega < 1$ is
allowed. Further, all the solitary waves in this case have a single hump.
In this case we find
\be\label{1.81}
Q = \frac{2(p+r)\sqrt{(p+r)}}{g^2 p r\sqrt{r (p+1)}} 
\tan^{-1}\left(\frac{\beta\sqrt{r+p}}{\omega\sqrt{r(p+1)}}\right)\,,
\ee
\be\label{1.82}
E = \frac{2(p+r)^2}{g^2 p r\sqrt{p r (p+1)(1-r)}} \tan^{-1}
\left(\frac{\sqrt{p(1-r)} \beta}{\sqrt{r(p+1)}}\right)\,.
\ee

\subsubsection{\bf $r > 1$}

In this case
\bea\label{1.83}
&&SS+VV+PP = \frac{1}{(\hat{g})^2} [SS +\frac{1}{\hat{p}} VV]\,, 
\nonumber \\
&&(\hat{g})^2 = \frac{r g^2}{(r-1)} < g^2\,, ~~\hat{p} 
= \frac{p(r-1)}{r+p}\,. 
\eea
Using the  results of the $SS+VV$ model \cite{fred7} we then predict
that in this case there is no restriction on $\omega$ and $0 < \omega < 1$ is
allowed. Further, all the solitary waves in this case have a double hump in case
\be\label{1.84}
0 <  \omega < \frac{p(r-2)-r}{2p(r-1)}\,,
\ee
while they have a single hump in case
\be\label{1.85}
\frac{p(r-2)- r}{2p(r-1)} < \omega < 1\,.
\ee

As an example if we let $p=3$, $r=5$, then the transition from a double hump to a 
single hump  takes place at $\omega=1/6$.  The behavior of $R^2(\omega)$ for this
value is shown in Fig. (\ref{rsqpp35}). 
\begin{figure}  
    \centering
    \includegraphics[width=0.7 \linewidth]{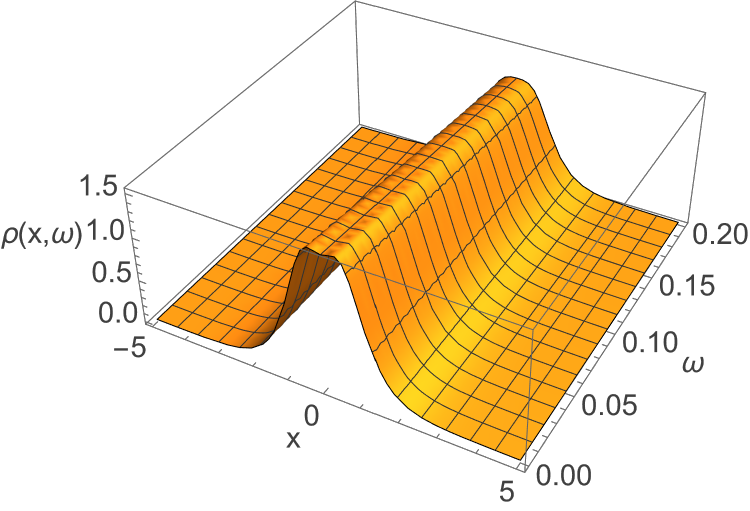}
    \caption{ \label{rsqpp35}  Plot of   $R^2(\omega) $    when  $\kappa=1, p=3,r=5$. }
\end{figure}
In this case we find
\be\label{1.86}
Q = \frac{2p(r-1)^2}{g^2 r\sqrt{r (p+1)(r+p)}} \tan^{-1}
\left(\frac{\sqrt{(r+p)} \beta}{\omega \sqrt{r(p+1)}}\right)\,,
\ee
\be\label{1.87}
E = \frac{2(r-1)\sqrt{p (r-1)}}{g^2 r\sqrt{(p+1)r}} \tanh^{-1}
\left(\frac{\sqrt{p(r-1)} \beta}{\sqrt{r(p+1)}}\right)\,.
\ee

\begin{figure}[t]
    \centering
    \includegraphics[width=0.7 \linewidth]{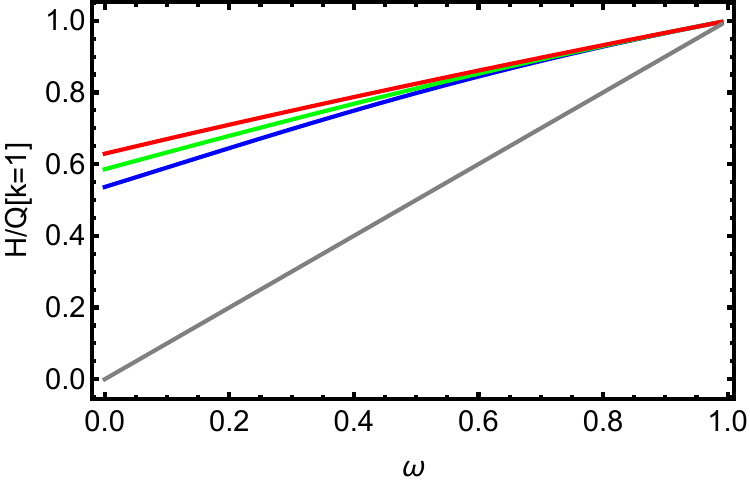}
    \caption{\label{EQk1}  Plot of $E(\omega,q,r)/Q(\omega,q,r)$ when  $\kappa=1$ for  $p=3$ and $r=1.1$ (red), 2 (green), 5 (blue) compared with $\omega$ (gray). }
\end{figure}

Keeping $p=3$ and letting $r=1.1,2,5$ we find that $E/Q (\omega)$ has the behavior shown in Fig. (\ref{EQk1}).
Since $E/Q(\omega) <1$ and the slope  of $E/Q (\omega) $ is positive  all the solitary waves are stable bound states in the non-relativistic regime.   Also we see that the relationship $E/Q > \omega$ is satisfied.

\subsection{Case IX: VV$-$SS$-$PP Interactions}

In this case $p=1, q < -1, r < 0$ so that at $\kappa = 1, R^2$ has a term
\be\label{1.88}
\frac{1}{g^2} \left[(1+\alpha^2 y^2)^2 -\frac{(1-\alpha^2 y^2)^2}{|q|} 
-\frac{4\alpha^2 y^2}{|r|}\right]\,,~~ |q| > 1, |r| > 0\,,
\ee
which can be re-expressed as
\be\label{1.89}
\frac{1}{g^2}\left[\frac{(|r|-1)}{|r|}(1+\alpha^2 y^2)^2 +\frac{(|q|-|r|)}{|q||r|} 
(1-\alpha^2 y^2)^2\right]\,.
\ee
Depending on the values of the parameters $|r| > 0, |q| > 1$ one gets
different combinations of SS and VV NLD interactions. 

\subsubsection{\bf $|r| = 1$}

In this case 
\be\label{1.90}
VV-SS-PP = \frac{1}{(\hat{g})^2} SS\,,~~(\hat{g})^2 = \frac{|q| g^2}{|q|-1} 
> g^2\,.
\ee
Using the  results of the SS model \cite{fred} we then predict
that in this case there is no restriction on $\omega$ and it can take any
value $0 < \omega < 1$. Further, all the solitary waves in this case have 
a double hump in case
\be\label{1.91}
0 < \omega < 1/2\,,
\ee
while they have a single hump in case
\be\label{1.92}
1/2 < \omega < 1\,.
\ee
Further, in this case we find
\be\label{1.93}
Q = \frac{2(|q|-1)\beta}{g^2 |q| \omega}\,,
\ee
\be\label{1.94}
E = \frac{4(|q|-1)}{g^2 |q|} \tanh^{-1}(\beta)\,.
\ee

\subsubsection{\bf $0 < |r| < 1$}

In this case
\bea\label{1.95}
&&VV-SS-PP = \frac{1}{(\hat{g})^2} [SS -\frac{1}{\hat{p}} VV]\,, 
\nonumber \\
&&(\hat{g})^2 = \frac{|q||r| g^2}{|q|-|r|} < g^2\,, ~~\hat{p} 
= \frac{(|q|-|r|)}{|q|(1-|r|)} > 1\,. 
\eea
Using the  results of the SS$-$VV model \cite{ak} we then predict
that in this case $\omega$ is restricted to
\be\label{1.96}
\sqrt{\frac{|q|(1-|r|)}{(|q|-|r|)}} \le \omega < 1\,.
\ee
Further, all the solitary waves in this case have a double hump in case
\be\label{1.97}
\sqrt{\frac{|q|(1-|r|)}{(|q|-|r|)}} \le \omega \le \frac{|q|(2-|r|)-|r|}
{2(|q|-|r|)}\,,
\ee
while they have a single hump in case
\be\label{1.98}
\frac{|q|(2-|r|)-|r|}{2(|q|-|r|)} < \omega < 1\,.
\ee
In this case we find
\be\label{1.99}
Q = \frac{2(|q|-|r|)^2 \sqrt{1-|r|}}{g^2 |r| \sqrt{|r| |q| (|q|-1)}} \tanh^{-1}
\left(\frac{\sqrt{|q|(1-|r|)} \beta}{\omega \sqrt{r(|q|-1)}}\right)\,,
\ee
\be\label{1.100}
E = \frac{2(|q|-|r|)\sqrt{|q|-|r|}}{g^2 |q| |r|\sqrt{|r|(|q|-1)}} \tanh^{-1}
\left(\frac{\sqrt{(|q|-|r|)} \beta}{\sqrt{|r|(|q|-1)}}\right)\,.
\ee

\subsubsection{\bf $|r| = |q| > 1$}

In this case
\be\label{1.101}
VV-SS-PP = \frac{1}{(\hat{g})^2} VV\,, ~~(\hat{g})^2 = \frac{|r| g^2}{|r|-1} > 
g^2\,. 
\ee
Using the  results of the $VV$ model \cite{fred} we then predict
that in this case there is no restriction on $\omega$ and $0 < \omega < 1$ is
allowed. Further, all the solitary waves in this case have a single hump.
In this case we find
\be\label{1.102}
Q = \frac{4(|r|-1)}{g^2 |r|} \tan^{-1}(\alpha)\,, ~~
E = \frac{2(|r|-1)\beta}{g^2 |r|}\,. 
\ee

\subsubsection{\bf $|r| > |q| > 1$}

In this case
\bea\label{1.103}
&&VV-SS-PP = \frac{1}{(\hat{g})^2} [VV -\frac{1}{|\hat{q}|} SS]\,, 
\nonumber \\
&&(\hat{g})^2 = \frac{|r| g^2}{|r|-1} > g^2\,, ~~\hat{q} 
= \frac{|q|(|r|-1)}{(|r| -|q|)} > 1\,. 
\eea
Using the  results of the $VV-SS$ model \cite{cgabs} we then predict
that in this case there is no restriction on $\omega$ and $0 < \omega < 1$ is
allowed. Further, all the solitary waves in this case have a single hump.

\begin{figure} 
    \centering
    \includegraphics[width=0.7 \linewidth]{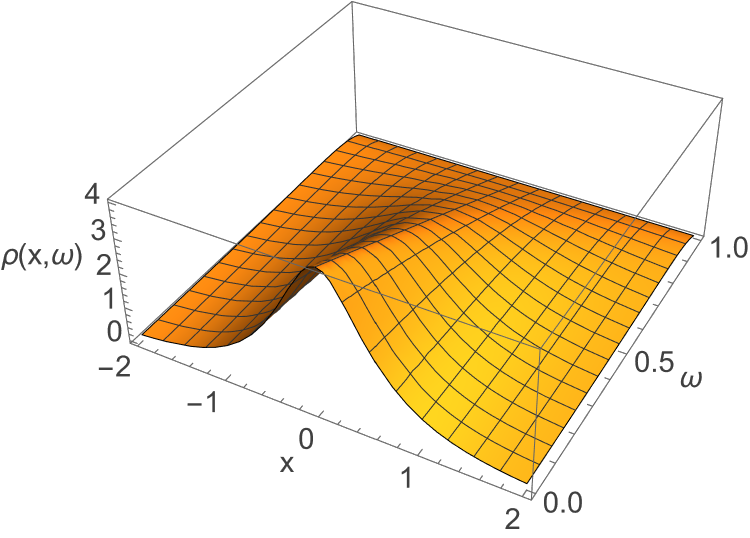}
    \caption{ \label{q-2r-3}  Plot of   $R^2(x,\omega) $    when  $\kappa=1, p=1,,q=-2,r=-3$.  }
\end{figure}

We display this case for $r=-3,q=-2$ in Fig. (\ref{q-2r-3}). 

Further, in this case we obtain
\be\label{1.104}
Q = \frac{2(|r|-1) \sqrt{|q|(|r|-1)}}{g^2 |r| \sqrt{|r|(|q|-|1|)}} 
\tan^{-1}\left(\frac{\sqrt{|q|(|r|-1)} \beta}{\omega \sqrt{[|r|(|q|-1)]}}\right)\,,
\ee
\be\label{1.105}
E = \frac{2 |q|(|r|-1)^2}{g^2 |r| \sqrt{|r|(|q|-1)(|r|-|q|)}} \tan^{-1}
\left(\frac{\sqrt{(|r|-|q|)} \beta}{\sqrt{|r|(|q|-1)}}\right)\,.
\ee

\subsubsection{\bf $|q| > |r| > 1$}

In this cse we get 
\bea\label{1.106}
&&VV-SS-PP = \frac{1}{(\hat{g})^2} [SS +\frac{1}{|\hat{p}|} VV]\,, 
\nonumber \\
&&(\hat{g})^2 = \frac{|r||q| g^2}{(|q|-|r|)} > g^2\,, ~~\hat{p} 
= \frac{(|q|-|r|)}{|q|(|r|-1)} > 0\,. 
\eea
Using the  results of the $SS+VV$ model \cite{fred7} we then predict
that in this case there is no restriction on $\omega$ and $0 < \omega < 1$ is
allowed. Further, all the solitary waves in this case have a double hump in case
\be\label{1.107}
0 < \omega \le \frac{|q|(2-|r|)-|r|}{2(|q|-|r|)}\,,
\ee
while they have a single hump in case
\be\label{1.108}
\frac{|q|(2-|r|)- |r|}{2(|q|-|r|)} < \omega < 1\,.
\ee
Further, in this case we get
\be\label{1.109}
Q = \frac{2(|q|-|r|)^2}{g^2 |q||r|\sqrt{|r||q|(|r|-1)(|q|-1)}} 
\tan^{-1}\left(\frac{\beta\sqrt{|q|(|r|-1)}}{\omega\sqrt{|r|(|q|-1)}}\right)\,,
\ee
\be\label{1.110}
E = \frac{2 (|q|-|r|)\sqrt{(|q|-|r|)}}{g^2 |q||r|\sqrt{|r|(|q|-1)}} 
\tan^{-1} \left(\frac{\beta\sqrt{|q|-|r}|}{\sqrt{(|q|-1)|r|}} \right)\,.
\ee

\subsection{\bf Case X: VV$-$PP}

In the limit of $|q| \rightarrow \infty$, the Case IX reduces to that of the
$VV-PP$ model. In particular, $VV-PP$ model at $\kappa = 1$ can be reexpressed 
as
\be\label{1.111}
\frac{1}{g^2}\left[\frac{(|r|-1)}{|r|}(1+\alpha^2 y^2)^2 +\frac{1}{|r|} 
(1-\alpha^2 y^2)^2\right]\,.
\ee
Now depending on the values of the parameters $|r| > 0$ one gets
different combinations of NLD interactions. We obtain

\subsubsection{\bf $|r| = 1$}

In this case we get 
\be\label{1.112}
VV-PP = \frac{1}{g^2} SS\,.
\ee
Using the  results of the SS model \cite{fred} we then predict
that in this case there is no restriction on $\omega$ and it can take any
value $0 < \omega < 1$. Further, all the solitary waves in this case have 
a double hump in case
\be\label{1.113}
0 < \omega < 1/2\,,
\ee
while they have a single hump in case
\be\label{1.114}
1/2 < \omega < 1\,.
\ee
Further, in this case we find
\be\label{1.115}
Q = \frac{2\beta}{g^2 \omega}\,,
\ee
\be\label{1.116}
E = \frac{4}{g^2} \tanh^{-1}(\beta)\,.
\ee

\subsubsection{\bf $0 < |r| < 1$}

In this case we get
\bea\label{1.117}
&&VV-PP = \frac{1}{(\hat{g})^2} [SS -\frac{1}{|\hat{p}|} VV]\,, 
\nonumber \\
&&(\hat{g})^2 = |r| g^2 < g^2\,, ~~\hat{p} 
= \frac{1}{(1-|r|)} > 1\,. 
\eea
Using the  results of the SS$-$VV model \cite{ak} we then predict
that in this case $\omega$ is restricted to
\be\label{1.118}
\sqrt{1-|r|} \le \omega < 1\,.
\ee
Further, all the solitary waves in this case have a double hump in case
\be\label{1.119}
\sqrt{1-|r|} \le \omega \le \frac{2-|r|}{2}\,,
\ee
while they have a single hump in case
\be\label{1.120}
\frac{(2-|r|)}{2} < \omega < 1\,.
\ee
In this case we find
\be\label{1.121}
Q = \frac{2}{g^2 |r| \sqrt{|r|}} \tanh^{-1}
\left(\frac{\sqrt{(1-|r|)} \beta}{\omega \sqrt{r}}\right)\,,
\ee
\be\label{1.122}
E = \frac{2}{g^2 |r|\sqrt{|r|}} \tanh^{-1}
\left(\frac{\beta}{\sqrt{|r|}}\right)\,.
\ee

\subsubsection{\bf $|r| > 1$}

In this case we get 
\bea\label{1.123}
&&VV-PP = \frac{1}{(\hat{g})^2} [SS +\frac{1}{|\hat{p}|} VV]\,, 
\nonumber \\
&&(\hat{g})^2 = |r| g^2 > g^2\,, ~~\hat{p} 
= \frac{1}{(|r|-1)} > 0\,. 
\eea
Using the  results of the $SS+VV$ model \cite{fred7} we then predict
that in this case there is no restriction on $\omega$ and $0 < \omega < 1$ is
allowed. Further, all the solitary waves in this case have a double hump in case
\be\label{1.124}
0 < \omega \le \frac{(2-|r|)}{2}\,,
\ee
while they have a single hump in case
\be\label{1.125}
\frac{(2-|r|)}{2} < \omega < 1\,.
\ee
Thus for $r \ge 2$ all the solitary waves have only a single hump.
In this case we get
\be\label{1.126}
Q = \frac{2}{g^2 |r|\sqrt{|r|(|r|-1)}} 
\tan^{-1}\left(\frac{\beta\sqrt{(|r|-1)}}{\omega\sqrt{|r|}}\right)\,,  
\ee
\be\label{1.127}
E = \frac{2}{g^2 |r|\sqrt{|r|}} 
\tanh^{-1} \left(\frac{\beta}{\sqrt{|r|}} \right)\,.
\ee

\section{Conclusions}

In this paper we have considered a three-parameter family of nonlinear Dirac 
equations at $\kappa=1$ with an arbitrary admixture of SS, VV and PP interactions.
We have shown that at $\kappa=1$ this model reduces to previously studied
 models having SS, VV, SS$-$VV,  VV$-$SS and VV+SS  interactions. 
 
We have considered the ten possible NLD models with various possible 
combinations of these three interactions and have shown that all the ten models
at $\kappa =1$ are equivalent to suitable combinations of SS and VV interactions 
which have been studied before \cite{fred, ak, cgabs, fred7}. Since it is well known 
that in all the models with combinations of SS and VV interactions all the solitary waves
at $\kappa = 1$ are solitary wave bound states, hence one can conclude that at 
$\kappa = 1$, all solitary waves in the NLD models with arbitrary mixture of 
SS, VV and PP interactions are also solitary wave bound states. We have shown 
that at $\kappa = 1$, two combinations of SS, VV and PP interactions are 
equivalent to $SS -(1/\hat{p}) VV$, while two other combinations of SS, VV and 
PP are equivalent to $VV-(1/\hat{q}) SS$. On the other hand, the remaining six 
combinations of SS, VV and PP are different combinations of SS and VV depending
on the values of the parameters $r, p, q$. We have considered these six models 
in some detail. We obtained a general expression for the minimum value of 
$\omega=\omega_{min}$ in order to have a finite charge $Q$.  We also obtained a
general expression for $\omega$ when there can be two-humped solitary wave 
solutions and also obtained the transition frequency $\omega_0$ above which
the solitary waves have a single hump. Finally, we also obtained expressions for
$Q$ and $E$ for all these ten models and have plotted some typical values of 
$E/Q( \omega)$ to show that all the solutions we have found are bound state solitary
waves with $E/(m Q) <1$ which are stable in the non-relativistic regime. 

This paper raises several questions some of which are:

\begin{enumerate}

\item We would like to know if the stability of solitary waves of the NLDE is 
linked with a single/double hump solution. So far the problem of demonstrating 
spectral stability  has proven to be very difficult \cite{Stathis}. Until this is resolved, 
interpreting numerical simulations of the evolution of single solitary waves or the 
scattering of two solitary waves would be difficult.  

\item There has been a lot of discussion in the literature about the behavior 
of the SS as well as the VV solitary waves in various external fields 
\cite{mertens, fred1}. It is then natural to examine the behavior of the solitary 
wave bound states in some of the NLD models with a suitable admixture 
of SS, VV and PP interactions under various external fields.

\item Finally, perhaps the most interesting question is if one can find 
relevance of some of the NLD models with an admixture of SS, VV and PP 
in the context of some physical phenomena, for example in the 
Bose-Einstein condensate or some other physical systems.

\item In this paper we have only considered the NLD models with an arbitrary 
admixture of SS, VV and PP interactions at $\kappa = 1$. The obvious question 
is: can one extend this to arbitrary nonlinearity, $\kappa$? In a companion paper we will 
extend our analysis to arbitrary integer $\kappa$. There one finds a major 
difference between the odd and the even integer values of $\kappa$. Whereas  
the even integer values of $\kappa$ lead to solitary waves that are not 
symmetric under parity, those with odd integral $\kappa$ lead to solitary waves
which are symmetric under parity.  One also finds that as one increases 
$\kappa$ one can not only have two-humped but even multi-humped solitary wave
solutions.

\end{enumerate}

\section{appendix: Vakhitov-Kolokolov (VK) Stability} 
For the NLSE,  linear stability is guaranteed by the VK stability criterion \cite{vk, sqm}.
 The VK criterion is that  given a solution of the NLSE, 
$\phi_\omega(x) e^{- i \omega t}$, it will be linearly stabile if
\be
\frac{dM}{d \omega} <0.
\ee
One lets 
\be
\psi(x,t) = [\phi_\omega(x) + r(x,t) )]e^{- i \omega t}
\ee 
and linearizes the equation for $r(x,t)$
\be
\partial_t(r(x,t)) = A_\omega r(x,t).
\ee
One then studies the eigenvalues of $A_\omega$. If the spectrum is imaginary then the solutions are spectrally stable. VK \cite{vk}  showed that when the spectrum is purely imaginary, $d M/d \omega <0.$   They also showed when $dM/ d \omega >0$ there is a real  positive eigenvalue so that there is a linear instability. For the NLSE one finds that when $\kappa >2$ the solitary waves are unstable. 

It is well known that VK criterion fails for the NLDE because even if $dQ/d \omega$  never vanishes, so that there are no bifurcations of nonzero real eigenvalues of the stability matrix from $\lambda = 0$, there still  may be nonzero real eigenvalues present in the spectrum of all solitary waves leading to linear instability  \cite{Comech1}.  So, one can only be certain that in the non-relativistic limit when $\omega/m \rightarrow 1$ the VK criterion for stability is valid.  For the  NLDE in $1+1$ dimensions, when the quantity $E/(mQ)$  is  less than one, the solitary wave is a bound state.  We  have shown here,  in \ref{QE} that its slope is opposite of the VK slope.  We noticed in previous papers on the NLDE with arbitrary nonlinearity parameter $\kappa$ that the slope of $E/Q(\omega) $ was positive up to $\kappa=2$ (where the NLSE became unstable), and when  $\kappa >2$ that the slope became negative in the non-relativistic regime.  Thus, it provides us confidence that up to $\kappa=2$ we expect the solitary waves we find are stable in the non-relativistic domain where the NLDE becomes the NLSE. 

\acknowledgments

We would like to thank Andrew Comech and Esthasios Charalampidis for useful 
discussions about stability. 
One of us (AK) is grateful to the Indian National Science Academy (INSA) for 
the award of the INSA Honorary Scientist position at Savitribai Phule Pune 
University, India. The work of AS was supported by the US Department of Energy. 

%
%

\end{document}